\documentclass[11pt,graphicx]{article}
\usepackage{amssymb,amsmath,amsfonts}
\usepackage{graphicx}
\usepackage{graphics}
\usepackage{eepic,epsfig}
\begin{document}
\title{Induced electrostatic self-interaction in the spacetime of a global monopole with inner structure}
\author{D. Barbosa$^2$ {\thanks{E-mail: denis.fisico@fisica.ufpb.br}}, J. Spinelly$^1$ {\thanks{E-mail: jspinelly@uepb.edu.br}}  
and E. R. Bezerra de Mello$^2$ \thanks{E-mail: emello@fisica.ufpb.br}\\
$^1$ Departamento de F\'{\i}sica-CCT, Universidade Estadual da Para\'{\i}ba\\
Juv\^encio Arruda S/N, C. Grande, PB, Brazil\\
$^2$ Departamento de F\'{\i}sica-CCEN, Universidade Federal da Para\'{\i}ba\\
58.059-970, C. Postal 5.008, J. Pessoa, PB,  Brazil}
\maketitle

\begin{abstract}
In this work we analise the electrostatic self-energy and self-force on a point-like electric charged particle induced by a global monopole spacetime considering a inner structure to it. In order to develop this analysis we calculate the three-dimensional Green function associated with this physical system. We explicitly show that for points inside and outside the monopole's core the self-energy presents two distinct contributions. The first is induced by the geometry associated with the spacetime under consideration, and the second one is a correction due to the non-vanishing inner structure attributed to it. Considering specifically the ballpoint-pen model for the region inside, we were able to obtain exact expressions for the self-energies in the regions outside and inside the monopole's core.
\\PACS numbers: $98.80.Cq$, $14.80.Hv$
\end{abstract}
\section{Introduction}
\label{Int}
Gravitational topological objects may have been formed in the early universe during its phase expansion due to spontaneously symmetry braking \cite{Kibble,V-S}. Depending on the topology of the vacuum manifold, ${\cal{M}}$, these are domain walls, strings, monopoles and texture, corresponding to the homotopy groups $\pi_0({\cal{M}})$, $\pi_1({\cal{M}})$, $\pi_2({\cal{M}})$ and $\pi_3({\cal{M}})$, respectively. Global monopoles are heavy topological objects formed in the phase transition of a system composed by a self-coupling iso-triplet scalar field, $\phi ^a$, whose original global $O(3)$ symmetry is spontaneously broken to $U(1)$. The scalar matter field plays the role of an order parameter which outside the monopole's core acquires a non-vanishing value. The global monopole was first introduced by Sokolov and Starobinsky in \cite{Soko}, and its gravitational effects have been analyzed by Barriola and Vilenkin \cite{BV}. In \cite{BV}, has been shown that for points far away from the monopole's center the geometry of the spacetime can be given by the following line element:
\begin{equation}
ds^2=-dt^2+dr^2+\alpha^2r^2(d\theta^2+\sin^2\theta d\varphi^2)\ ,  
\label{gm}
\end{equation}
where the parameter $\alpha^2$, smaller than unity, depends on the energy scale $\eta$ where the global symmetry is spontaneously broken. The spacetime described by (\ref{gm}) has a non-vanishing scalar curvature, $R=\frac{2(1-\alpha^2)}{\alpha^2r^2}$, and the solid angle of a sphere of unit radius is smaller than the usual one, presenting in this way, a solid angle deficit: $\delta\Omega=4\pi^2(1-\alpha^2)$. Although the geometric properties of the spacetime outside the monopole are very well understood, the analysis of the metric tensor in the region inside requires the complete knowledge of the energy-momentum tensor associated with the scalar field, $\phi ^{a}$, which on the other hand depends on the knowledge of the components of the metric tensor, providing, in this way, a non solvable integral equation \cite{Mello1}. Because of this fact, many interesting investigations of physical effects associated with global monopole consider this object as a point-like defect. In this way, it is considered that the geometry of the whole spacetime is described by (\ref{gm}); however, adopting this model, calculations of vacuum polarization effects, for example, present divergences on the monopole's core. \footnote{Vacuum polarization effects associated with bosonic and fermionic quantum fields, have been analyzed in \cite{Lousto} and \cite{Mello2}, respectively, considering the global monopole as point-like defect.} 

One of most interesting phenomenon associated with the gravitational topological defects, is related with the induced self-energy on an electric charged particle placed at rest in their neighborhood. This effect has been analysed in an idealized cosmic string spacetime by Linet \cite{Linet} and Smith \cite{Smith}, independently, and in the spacetime of a point-like global monopole in \cite{Mello3}. In these analysis, the induced self-energies present divergences on the respective defects' core. In order to avoid this problem, a more realistic model for the defects should take into account a inner structure for them. As to cosmic string, two different models have been adopted to describe the geometry inside it: the ballpoint-pen model proposed independently by Gott and Hiscock \cite{Gott}, replaces the conical singularity at the string axis by a constant curvature spacetime in the interior region, and flower-pot model \cite{BA}, presents the curvature concentrated on a ring with the spacetime inside the string been flat. Khusnutdinov and Bezerra in \cite{NV}, revisited the induced electrostatic self-energy problem considering the Hiscock and Gott model for the region inside the string. Recently the electrostatic self-energy problem in the context of global monopole has been analyzed considering the flower-pot model \cite{Mello4} for the region inside.\footnote{The analysis of vacuum polarization effect associated with scalar and fermionic fields, considering the flower-pot model for the region inside the monopole, have been developed in \cite{Mello5} and \cite{Mello6}, respectively} In order to complete the analysis of electrostatic self-energy in the global monopole spacetime, the propose of this paper is to consider the ballpoint-pen model for the region inside. 

This paper is organized as follows: In section \ref{Sect1} is presented the model adopted to describe the spacetime in the region inside the global monopole and some its geometric properties. In section \ref{Sect2}, we write the Maxwell equations in the corresponding spacetime, and derive the Green function for points outside and inside to the monopole's core. In section \ref{Sect3} we calculate the corresponding renormalized induced electrostatic self-energy, and give its behavior in various asymptotic regions of the parameters. In section \ref{Sect4} we present our conclusions and more relevant remarks.

\section{The model}
\label{Sect1}
The simplest model which gives rise to a global monopole has been proposed by Barriola and Vilenkin \cite{BV}, and is described by a Lagrangian whose original $O(3)$ global gauge symmetry is spontaneously broken to $U(1)$. The influence of this object on the geometry of the spacetime can be analysed by coupling the energy-momentum tensor associated with this matter field with Einstein equations. Barriola and Vilenkin have shown that for points very far away form the monopole's center, the spacetime is described by the line element (\ref{gm}). Although there are no analytical solutions for the metric tensors in the inner region, Harari and Loust\'o \cite{HL} presented a simplified version for the geometry in this region. They have shown that it can be represented by a de Sitter spacetime, whose constant scalar curvature is proportional to the energy scale $\eta$ where the symmetry is broken. Although this model shares the main features of the exact solution, the analysis of the three-dimensional Green function in the region inside the monopole is a very difficult task.\footnote{The four-dimensional massive scalar Green function in de Sitter spacetime is investigated in \cite{Cand} (see also \cite{BD}).} Another model which also shares some features with the exact solution and allows us to obtain closed solutions for this function is the ballpoint-pen model. This model is given as shown below: 
\begin{itemize}
\item For the region outside the monopole the metric of the spacetime is given by (\ref{gm}), where the radial coordinates is defined in the interval $[r_0, \ \infty)$, being $r_0$ the radius of the monopole.
\item For inner region the metric tensor is given by the following line element
\begin{eqnarray}
\label{gm1}
	ds^2=-dt^2+d\rho^2+\left(\frac{\rho_0}\epsilon\right)^2\sin^2\left(\frac{\epsilon\rho}{\rho_0}\right)(d\theta^2+\sin^2\theta d\varphi^2) \ ,
\end{eqnarray}
with the new radial coordinate defined in the interval $[0, \ \rho_0]$. 
\end{itemize}
The inner solution can match the exterior solution without the necessity to include an infinitely thin shell at the boundary, provided the junction conditions below:
\begin{eqnarray}
\alpha r_0=\frac{\rho_0}\epsilon\sin\epsilon \ , \ {\rm with} \ \ \alpha=\cos\epsilon \ .
\end{eqnarray}

In the inner region, the non zero components of the Riemann and Ricci tensors and scalar curvature are:
\begin{eqnarray}
	R^{\rho\theta}_{\rho\theta}=R^{\rho\varphi}_{\rho\varphi}=R^{\varphi\theta}_{\varphi\theta}=\frac{\epsilon^2}{\rho_0^2} \ , \ R^\rho_\rho=R_\theta^\theta =R^\varphi_\varphi=2\frac{\epsilon^2}{\rho_0^2} \ , \ {\rm and} \ R=6\frac{\epsilon^2}{\rho_0^2} \ .
\end{eqnarray}

It is possible to describe the metric tensor in the inner region by using the radial $r$ coordinate through the relation
\begin{eqnarray}
	\sin\left(\frac{\epsilon\rho}{\rho_0}\right)=\frac r{r_0}\sin\epsilon \ .
\label{rela}
\end{eqnarray}
In this case, the line element (\ref{gm1}) can be written by
\begin{eqnarray}
\label{gm2}
	ds^{2}=-dt^{2}+P^2(r)dr^{2}+\alpha^{2}r^{2}(d\theta^{2}+\sin^{2}\theta d\varphi^{2}) \ ,
\end{eqnarray}
where
\begin{equation}
P(r)=\left\{
\begin{array}{ll}
\frac\alpha{\sqrt{1-\frac{r^2}{r_0^2}\sin^2\epsilon}}  & \mathrm{for}\ r\leq r_0 \ \\
1 & \mathrm{for}\ r\geq r_0 
\end{array}
.\right.  
\end{equation}

\section{Green function}
\label{Sect2}
With the objective to construct the Green function associated with an electric charged particle at rest in the spacetime of a global monopole, we write down the Maxwell equation in an arbitrary curved spacetime
\begin{eqnarray}
\label{Meq}
\Box A^\mu+R^\mu_\nu A^\nu=-4\pi j^\mu \ ,
\end{eqnarray}
with
\begin{eqnarray}
\Box A^{\mu}&=&g^{\alpha\beta}(\partial_\alpha\partial_\beta A^{\mu})+g^{\alpha\beta}(\partial_\alpha\Gamma_{\beta\gamma}^{\mu})A^{\gamma}+g^{\alpha\beta}\Gamma_{\beta\gamma}^{\mu}(\partial_\alpha A^{\gamma})+g^{\alpha\beta}\Gamma_{\alpha\nu}^{\mu}(\partial_\beta A^{\nu})\nonumber\\
&+&g^{\alpha\beta}\Gamma_{\alpha\nu}^{\mu}\Gamma_{\beta\gamma}^{\nu}A^{\gamma}-g^{\alpha\beta}\Gamma_{\alpha\beta}^{\rho}(\partial_\rho A^{\mu})-g^{\alpha\beta}\Gamma_{\alpha\beta}^{\rho}\Gamma_{\rho\gamma}^{\mu}A^{\gamma} \ ,
\end{eqnarray}
where $A^\mu$ and $j^\mu$ are the four-vector potential and current, respectively.  For a point-like particle at rest with coordinates $\vec{r'}=(r',\theta',\varphi')$, the static four-vector current and potential read: $j^\mu=(j^0,0,0,0)$ and $A^\mu=(A^0,0,0,0)$. The only nontrivial component of (\ref{Meq}) is for $\mu=0$, with
\begin{equation}
j^{0}(x)=q\frac{\delta (\vec{r}-\vec{r}')}{\sqrt{-g}}\ ,  
\label{J0}
\end{equation}%
being $q$ is the charge of the particle. The corresponding Maxwell equation is written as
\begin{eqnarray}
	\Delta A^{0}=-4\pi j^{0} \ ,
\end{eqnarray}
where the three-dimensional Laplacian operator in coordinates system given in (\ref{gm2}) is
\begin{eqnarray}
\label{A}
\Delta A^{0}=\left[\frac{1}{r^{2}P(r)}\frac{\partial}{\partial r}\left(\frac{r^{2}}{P(r)}\frac{\partial}{\partial r}\right)- \frac{\vec{L}^{2}}{\alpha^{2}r^{2}}\right]A^{0} \ ,
\end{eqnarray}
being $\vec{L}$ the angular momentum operator. Moreover in this coordinate $\sqrt{-g}=P(r)\sin{\theta}\alpha^{2}r^{2}$. 

The Green function associated with the respective operator is obtained by substituting
\begin{equation}
A^0(\vec{r})=4\pi qG(\vec{r},\vec{r}') \ 
\label{A0}
\end{equation}
into the above equations. So, this non-homogeneous differential equation reads:
\begin{eqnarray}
\label{Green-a}
\left[\frac{\partial}{\partial r}\left(\frac{r^{2}}{P(r)}\frac{\partial}{\partial r}\right)-\frac{P(r)}{\alpha^2}{\vec{L}^{2}}\right] G(\vec{r},\vec{r}')=-\frac{\delta(r-r')\delta(\theta-\theta')\delta(\varphi-\varphi')}{\alpha^2\sin\theta} \ .
\end{eqnarray}

Taking into account the spherical symmetry of the problem, we may present the Green function as the expansion
\begin{equation}
G(\vec{r},\vec{r}')=\sum_{l=0}^\infty\sum_{m=-l}^lg_l(r,r')Y_l^m(\theta ,\varphi )Y_l^{m\ast}(\theta',\varphi')\ ,  \label{Green-b}
\end{equation}
with $Y_l^m(\theta ,\varphi )$ being the ordinary spherical harmonics. Substituting (\ref{Green-b}) into (\ref{Green-a}) and using the well known
closure relation for the spherical harmonics, we arrive at the following differential equation for the radial function:
\begin{eqnarray}
\label{gr}
\left[\frac{d}{dr}\left(\frac{r^2}{P(r)}\frac{d}{dr}\right)-\frac{P(r)l(l+1)}{\alpha^2}\right]g_l(r,r')= -\frac{\delta(r-r')}{\alpha^{2}} \ .
\end{eqnarray}
The junction of the first radial derivative at $r=r'$ is obtained by integrating the above equation about this point:
\begin{equation}
\label{f1}
\frac{dg_l(r,r')}{dr}|_{r=r^{'+}}-\frac{dg_l(r,r')}{dr}|_{r=r^{'-}}=-\frac{P(r')}{\alpha^2r'^2} \ .
\end{equation}
The radial Green function is calculated by the standard method:
\begin{eqnarray}
	g_l(r,r')=\Theta(r'-r)R_{1l}(r)R_{2l}(r')+\Theta(r-r')R_{1l}(r')R_{2l}(r) \ ,
\end{eqnarray}
where $R_{1l}(r)$ and $R_{2l}(r)$ are the two linearly independent solutions of the homogeneous equation corresponding to (\ref{gr}). We  assume that $R_{1l}(r)$ is regular at the core center, $R_{2l}(r)$ goes to zero at infinity, and that these solutions are normalized by the Wronskian relation
\begin{equation}
R_{1l}(r)R_{2l}^{\prime }(r)-R_{1l}^{\prime }(r)R_{2l}(r)=-\frac{P(r)}{\alpha^2r^2} \ .  \label{Wronin}
\end{equation}Moreover, both solutions must be regular with their first radial derivative regular at the boundary, i.e., 
\begin{eqnarray}
\label{f2}
	R_{jl}(r)|_{r=r_0-}&=&R_{jl}(r)|_{r=r_0+} \ , \nonumber\\
	R'_{jl}(r)|_{r=r_0-}&=&R'_{jl}(r)|_{r=r_0+} \ ,
\end{eqnarray}
for $j=1, \ 2$.

In the region outside the core the linearly independent solutions to the corresponding homogeneous equation are the functions $r^{\lambda _l}$ and $
r^{-1-\lambda_l}$, where
\begin{equation}
\lambda_l=-\frac12+\frac1{2\alpha}\sqrt{\alpha^2+4l(l+1)}\geq 0 \ ,  \label{lam}
\end{equation}
and for the region inside, in coordinates given in (\ref{gm2}), the regular and singular solutions are $\frac{P^{-l-1/2}_{1/2}(x_r)}{\sqrt{r}}$ and $\frac{P^{l+1/2}_{1/2}(x_r)}{\sqrt{r}}$, respectively, with $x_r=\alpha(P(r))^{-1}$, being $P_\mu^\nu(x)$ the associated Legendre functions. 

Now taking into account (\ref{f1}), (\ref{Wronin}) and (\ref{f2}), the radial Green functions are given by:
\begin{itemize}
\item For the charged particle outside the monopole's core, $r'> \ r_0$, we have:
\begin{eqnarray}
\label{g1}	g_1(r,r')=\frac1{\alpha^2(2\lambda_l+1)}\frac{(1+\gamma_l(\alpha))}{\sqrt{rr_0}}\left(\frac{r_{0}}{r^{'}}\right)^{\lambda_l+1}\frac{P_{1/2}^{-l-1/2} (x_r)}{P_{1/2}^{-l-1/2}(\alpha)} \ , \ {\rm for} \ \ r\leq r_0 
\end{eqnarray}
and
\begin{eqnarray}
\label{g2}	g_{l}(r,r')=\frac1{\alpha^2(2\lambda_l+1)}\left[\left(\frac{r_{<}}{r_{>}}\right)^{\lambda_l}\frac1{r_{>}}+\gamma_l(\alpha)r_0^{\lambda_l}\left(\frac{r_{0}}{rr'}\right)^{1+\lambda_l}\right] \ , \ {\rm for} \ \ r\geq r_0 \ 
\end{eqnarray}
with 
\begin{eqnarray}
	\gamma_l(\alpha)=\frac{\alpha\lambda_lP_{1/2}^{-l-1/2}(\alpha)-lP_{-1/2}^{-l-1/2}(\alpha)}{\alpha(\lambda_l+1)P_{1/2}^{-l-1/2}(\alpha)+lP_{-1/2}^{-l- 1/2} (\alpha)} \ .
	\label{gamma}
\end{eqnarray}
\item For the charged particle inside the monopole's core, $r'< \ r_0$, we have:
\begin{eqnarray}
\label{g3}
g_l(r,r')=\frac1{\alpha\kappa_l{\sqrt{rr'}}} P_{1/2}^{-l-1/2}(x_{r_<})\left[P_{1/2}^{-l-1/2}(x_{r_>})\eta_l(\alpha)+P_{1/2}^{l+1/2}(x_{r_>})\right] \ , \ {\rm for} \ \ r\leq r_0
\end{eqnarray}
and
\begin{eqnarray}
\label{g4}
g_l(r,r')=\frac1{\alpha\kappa_l{\sqrt{rr'}}}\left(\frac{r_0}r\right)^{\lambda_l +1}P_{1/2}^{-l-1/2}(x_{r'}) \left[P_{1/2}^{-l-1/2}(\alpha)\eta_l(\alpha)+P_{1/2}^{l+1/2}(\alpha)\right] \ , \ {\rm for} \ \ r\geq r_0 \ 
\end{eqnarray}
with \footnote{For the derivation of (\ref{g3}) and (\ref{g4}), we have used the corresponding Wronskian for the associated Legendre functions \cite{Grad}: ${\cal{W}}[P_\nu^{-\mu}(x), \ P_\nu^\mu(x)]=\frac{2\sin\mu\pi}{\pi(1-x^2)}$.}
\begin{eqnarray}
\eta_l(\alpha)=\frac{(l+1)P_{-1/2}^{l+1/2}(\alpha)-\alpha(\lambda_l+1)P_{1/2}^{l+1/2}(\alpha)}{lP_{-1/2}^{-l-1/2}(\alpha)+\alpha(\lambda_l+1) P_{1/2}^{-l-1/2}(\alpha)} \ ,
\end{eqnarray}
and
\begin{eqnarray}
\kappa_l=\frac2\pi(-1)^l \ .
\end{eqnarray}
\end{itemize}
In these formulas we have adopted $r_<=\min (r,r')$ and $r_>=\max (r,r')$, with the same convention for $x_{r_<}$ and $x_{r_>}$.

First, let us consider the case where the charged particle is outside the monopole's core. Substituting (\ref{g2}) into (\ref{Green-b}), we have
\begin{eqnarray}
\label{ap}
G_{out}(r,r')&=&\frac1{4\pi\alpha^2r_>}\sum_{l=0}^\infty\frac{2l+1}{2\lambda_l+1}\left(\frac{r_<}{r_>}\right)^{\lambda_l}P_{l}(\cos\gamma)\nonumber\\
&+&\frac{1}{4\pi\alpha^2r_0}\sum_{l=0}^\infty\frac{2l+1}{2\lambda_l+1}\gamma_l(\alpha)\left(\frac{r_0^2}{rr'}\right)^{1+\lambda_l}P_l(\cos\gamma) \ .
\end{eqnarray}
where the first term on the right hand side of (\ref{ap}) is the Green function for the geometry of a point-like global monopole, $G_m(r,r')$, and the second, $G_c(r,r')$, is induced by the non-trivial structure of the core. $\gamma$ is the angle between the directions $(\theta,\phi)$ and $(\theta',\phi')$.

For the particle inside the monopole's core, the Green function is given by substituting (\ref{g3}) into (\ref{Green-b}), the result is:
\begin{eqnarray}
\label{bl}
G(\vec{r},\vec{r}')&=&\frac1{4\pi\alpha\sqrt{r'r}}\sum_{l=0}^\infty\frac{2l+1}{\kappa_l}P_{\frac12}^{l+\frac12}\left(x_{r_>}\right)P_{\frac12}^{-l-\frac12}\left(x_{r_<}\right)P_l(\cos{\gamma})\nonumber\\
&+&\frac1{4\pi\alpha\sqrt{r'r}}\sum_{l=0}^\infty\frac{2l+1}{\kappa_l}P_{\frac12}^{-l-\frac12}\left(x_{r_>}\right)P_{\frac12}^{-l-\frac12}\left(x_{r_<}\right)P_l(\cos{\gamma})\eta(\alpha) \ ,
\end{eqnarray}
being 
\begin{eqnarray}
\label{X}
	x_{r_>}=\sqrt{1-\frac{r_{>}^{2}\sin^{2}\epsilon}{r_{0}^{2}}} \ , \  {\rm and} \  x_{r_<}=\sqrt{1-\frac{r_{<}^{2}\sin^{2}\epsilon}{r_{0}^{2}}} \ .
\end{eqnarray}
Also, this function presents two contributions: The first one, $G_0(\vec{r},\vec{r}')$, is the Green function in the background geometry described by the line element (\ref{gm1}) using the external radial coordinate, and the second, $G_{\alpha}(\vec{r},\vec{r}')$, is due to the geometry of global monopole for $r>r_0$.

\section{Self-energy}
\label{Sect3}
According to \cite{Linet,Smith}, the induced electrostatic self-energy associated with the electric charge is given by
\begin{equation}
U_{Ele}(\vec{r})=2\pi q^2\lim_{\vec{r}'\rightarrow \vec{r}}G_{Ren}(\vec{r}',\vec{r}) \ ,
\label{SE}
\end{equation}
where $G_{Ren}$ is the renormalized Green function defined as
\begin{eqnarray}
	G_{Ren}(\vec{r}',\vec{r})=G(\vec{r}',\vec{r})-G_H(\vec{r}',\vec{r}) \ , 
\end{eqnarray}
being $G_H$ the Hadamard function. 

The general expression for the Hadamard function in a three-dimensional space is
\begin{eqnarray}
	G_H(x',x)=\frac{\Delta^{1/2}(x',x)}{4\pi}\frac1{\sqrt{2\sigma(x',x)}} \ , 
\end{eqnarray}
where $\Delta$, the Van Vleck-Morette determinant, is given by
\begin{eqnarray}
	\Delta^{1/2}=1+\frac1{12}R_{ij}\sigma^{i}\sigma^j+...
\end{eqnarray}
and $2\sigma(x',x)$ the geodesic interval between $x'$ and $x$.

Because the self-energy depends only on the radial coordinates, the respective induced self-force is given by
\begin{eqnarray}
\label{SFO}
	{\vec{F}}_{Ele}(r)=-\frac d{dr}U_{Ele}(r){\hat{r}} \ .
\end{eqnarray}

In the rest of this section we shall calculate the induced electrostatic self-energy for the two cases specified: $i)$ for the charge outside the monopole, and $ii)$ for the charge inside it.

\subsection{Self-energy in the region outside the monopole}
As we have seen, for the region outside the monopole's core the Green function to be considered in the calculation of the self-energy is given by (\ref{ap}). According to (\ref{SE}), the induced electrostatic self-energy is given by taking the coincidence limit of the renormalized Green function.  We observe that for points with $r>r_0$, the core induced part of this function is finite in the coincidence limit, and that the divergent contribution appears only in the point-like part only. So, in the renormalization procedure the only part to be considered is $G_m$:
\begin{eqnarray}
	G_{m,ren}(\vec{r},\vec{r})=\lim_{{\vec{r}}\to\vec{r}}\left[G_m(\vec{r}',\vec{r})-G_H(\vec{r}',\vec{r})\right] \ .
\end{eqnarray}
 
In the coincidence limit, let us first take $\gamma=0$, so, for points along the radial distance, the singular part of the Hadamard function reads,
\begin{eqnarray}
	G_H(r',r)=\frac1{4\pi|r'-r|} \ .
\end{eqnarray}
Now, by using $G_m(r',r)$ given in (\ref{ap}), we have:
\begin{equation}
G_{{m,ren}}(r,r)=\frac1{4\pi r_{0}}\lim_{t\to1}\left[\frac1{\alpha}\sum_{l=0}^\infty\frac{2l+1}{\sqrt{\alpha
^{2}+4l(l+1)}}\ t^{\lambda_l}-\frac{1}{1-t}\right] ,  \label{Gmren1}
\end{equation}
where $t=r_{<}/r_{>}$. In order to evaluate the limit on the right hand side of the above equation, we note that
\begin{equation}
\lim_{t\to 1}\left( \frac{1}{\alpha }\sum_{l=0}^{\infty }t^{l/\alpha+1/2\alpha -1/2}-\frac{1}{1-t}\right) =0.  
\label{RelLim}
\end{equation}
So, on the basis of this relation, replacing in (\ref{Gmren1}) $1/(1-t)$ by the first term in the brackets in (\ref{RelLim}), we find
\begin{equation}
G_{{m,ren}}(r,r)=\frac{S(\alpha )}{4\pi\alpha r} \ .
\label{Gmren2}
\end{equation}
In the above expression we have introduced the notation
\begin{equation}
S(\alpha )=\sum_{l=0}^{\infty }\left[\frac{2l+1}{\sqrt{\alpha ^{2}+4l(l+1)}}-1\right] \ .  
\label{falfa}
\end{equation}

The function $S(\alpha )$ is positive (negative) for $\alpha <1$ ($\alpha >1$) and, hence, the corresponding self-force is repulsive (attractive).
Developing a series expansion in the parameter $\eta ^{2}_0=1-\alpha ^{2}$, we can see that
\begin{equation}
S(\alpha )=\sum_{n=1}^{\infty }\frac{(\pi \eta_0 )^{2n}}{2(n!)^{2}}|B_{2n}|(1-2^{-2n}) \ ,
\end{equation}
where $B_{n}$ are the Bernoulli numbers. The leading term in the expression on the right is $\pi^2 (1-\alpha ^{2})/16$. For large values $\alpha $ the
main contribution into the series in (\ref{falfa}) comes from large values $l$. Replacing the summation by the integration we can see that in the limit $\alpha \rightarrow \infty $ the function $S(\alpha )$ tends to the limiting value $-1/2$. For small values $\alpha $, $\alpha \ll 1$, the main
contribution comes from the term $l=0$ and one has $S(\alpha )\approx1/\alpha ^{2}$.

\begin{figure}[tbph]
\begin{center}
\epsfig{figure=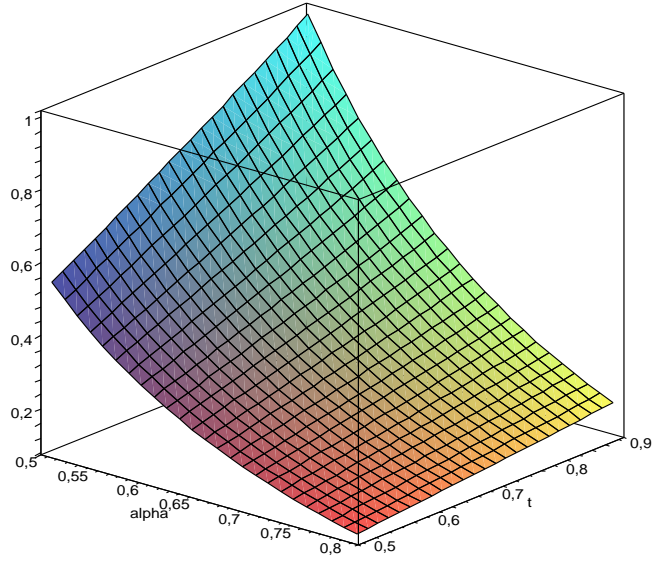,width=10.cm,height=9.cm}
\end{center}
\caption{Electrostatic self-energy for a charge outside the monopole core as a function of the monopole parameter $\alpha $ and rescaled radial coordinate $t_{out}=r_0/r$.}
\label{fig1}
\end{figure}

Now combining formulas (\ref{SE}), (\ref{ap}) and (\ref{Gmren2}), the induced electrostatic self-energy reads
\begin{eqnarray}
	U_{Ele}(r)=\frac{q^2}{2\alpha r}S(\alpha)+\frac{q^2}{2\alpha r} \sum_{l=0}^\infty\frac{\gamma_l(\alpha)(2l+1)} {\sqrt{\alpha^2+4l(l+1)}}\left(\frac{r_0}{r}\right)^{1+2\lambda_l} \ .
	\label{SE1}
\end{eqnarray}
The second term on the right hand side of the above expression if positive for $\alpha< \ 1$ and negative for $\alpha> \ 1$, consequently, according to the properties discussed for the function $S(\alpha)$, the corresponding self-force given by
\begin{eqnarray}
	{\vec{F}}_{Ele}(r)=\frac{q^2S(\alpha)}{2\alpha r^2}{\hat{r}}+\frac{q^2}{\alpha r^2} \sum_{l=0}^\infty\frac{\gamma_l(\alpha)(2l+1)(\lambda_l+1)} {\sqrt{\alpha^2+4l(l+1)}}\left(\frac{r_0}{r}\right)^{1+2\lambda_l}{\hat{r}} \ , 
\end{eqnarray}
is repulsive for the first case and attractive for the second one. For large distance from the monopole's core, the main contribution associated with the induced core part  comes from the $l=1$ component (the $l=0$ component vanishes), so we have
\begin{eqnarray}
U_{Ele}(r)\approx\frac{q^2}{2\alpha r}\left[S(\alpha)+\frac{3\gamma_1(\alpha)}{\sqrt{\alpha^2+8}}\left(\frac{r_0}r\right)^{\sqrt{1+8/\alpha^2}}\right] \ .
\end{eqnarray}
There is a suppressed factor, $\left(\frac{r_0}r\right)^{\sqrt{1+8/\alpha^2}}$, in the core-induced contribution. The core-induced part is divergent at the boundary, $r=r_0$. In order to verify this singular behavior it is sufficient to analyse this quantity for large values of $l$. The asymptotic expression for (\ref{gamma}) can be obtained by using the relation between Legendre function and hypergeometric one \cite{Grad},
\begin{eqnarray}
\label{Legend}
	P_\nu^\mu(x)=\frac1{\Gamma(1-\mu)}\left(\frac{1+x}{1-x}\right)^{\frac\mu2}F\left(-\nu, \ \nu+1; \ 1-\mu; \frac{1-x}2\right) \ .
\end{eqnarray}
For the case under consideration $\nu=\pm\frac12$ and $\mu=-l-\frac12$. For large value of $l$ the leading term of the corresponding hypergeometric function is the unity (see \cite{Abra}), and we can write
\begin{eqnarray}
	\gamma_l(\alpha)\approx-\frac1{4l}(\alpha-1)+O\left(\frac1{l^2}\right) \ .
\end{eqnarray}
On the other hand $\frac{2l+1}{\sqrt{\alpha^2+4l(l+1)}}\approx1$. So after some additional steps we find
\begin{eqnarray}
U_{Ele}(r)\approx\frac{q^2}{8\alpha r_0}(\alpha-1)\ln\left(1-(r_0/r)^{1/\alpha}\right)	\ ,
\end{eqnarray}
and we can see that the electrostatic self-energy is dominated by the core-induced part. The analysis of the behavior of the complete self-interaction can only be provided numerically. So, in in figure \ref{fig1} we have plotted $r_0U_{Ele}/q^2$ as function of the parameter $\alpha $ and $r_0/r$.

\subsection{Self-energy in the region inside the monopole}
In order to develop the analysis of the self-energy in the region inside the monopole, we should consider the corresponding Green function given in (\ref{bl}). As in the previous analysis, the core-induced contribution of this function is finite at the coincidence limit for $r<r_0$. The only divergent contribution in this limit comes from $G_0(\vec{r}',\vec{r})$. The Hadmard function, in the region inside the monopole, needed to renormalize this Green function can be given in a simple way by using the coordinate system defined in (\ref{gm1}). Taking the coincidence limit in the angular variables first, this function reads:
\begin{eqnarray}
\label{Hada}
	G_H(\rho',\rho)=\frac1{4\pi}\frac1{|\rho'-\rho|} \ .
\end{eqnarray}
Because we have obtained the Green function (\ref{bl}) in the coordinate system defined in (\ref{gm2}), it is more convenient to express the above Hadamard function in this system. So by using (\ref{rela}) the singular behavior of (\ref{Hada}) is
\begin{eqnarray}
	G_H(r',r)=\frac1{4\pi\alpha}\frac{\sqrt{1-(r/r_0)^2\sin^2\epsilon}}{|r'-r|} \ .
\end{eqnarray}
Finally the renormalized Green function is given by
\begin{eqnarray}
	G_{0,ren}(r,r)=\lim_{r'\to r}\left[G_0(r',r)-G_H(r',r)\right] \ .
\end{eqnarray}	
So developing some intermediate steps, we obtain the following expression for $G_{0,ren}(r',r)$:\footnote{In appendix \ref{app} we present the explicit calculation adopted to obtain (\ref{Gr0}).}
\begin{eqnarray}
\label{Gr0}
G_{0,m}(r,r)=\frac1{4\pi\alpha r}{\bar{S}}(\alpha,r/r_0) \ ,
\end{eqnarray}
where
\begin{eqnarray}
\label{Gr1}
{\bar{S}}(\alpha,r/r_0)=\sum_{l=0}^\infty\left[F\left( -\frac12, \ \frac32; \ \frac12-l; \ \frac{1-x_{r}}2\right)F\left( -\frac12, \ \frac32; \ \frac32+l; \ \frac{1-x_{r}}2\right)-1\right] 
\end{eqnarray}
and
\begin{eqnarray}
\label{Gr2}
	x_r=\sqrt{1-(r/r_0)^2\sin^2\epsilon} \ .
\end{eqnarray}

Now we are in position to write the electrostatic self-energy for a charged particle inside the monopole's core:
\begin{eqnarray}
\label{ELE}
	U_{Ele}(r)=\frac{q^2}{2\alpha r}{\bar{S}}(\alpha,r/r_0)+\frac{q^2}{4\alpha r}\sum_{l=0}^\infty(2l+1)\pi(-1)^l\eta_l(\alpha)(P_{\frac12}^{-l- \frac l2}(x_r))^2 \ .
\end{eqnarray}

As in the case of exterior region, the self-energy is positive for $\alpha<1$ and negative for $\alpha>1$. The corresponding self-force is obtained by using (\ref{SFO}), and is repulsive with respect to the boundary of the monopole core in the first case and attractive in the second case. Near the core's center the most relevant contribution comes from the $l=0$ component of the core-induced contribution. In this region the self-energy behaves as
\begin{eqnarray}
	U_{Ele}(r)\approx\frac{q^2}{2\alpha^2 r_0}\eta_0(\alpha)(1-\alpha^2) \ .
\end{eqnarray}
Also the core-induced term diverges at the at the boundary. In order to see that let us analyse this terms for large value of $l$. In this limit we can verify that
\begin{eqnarray}
	\eta_l(\alpha)\approx\frac{1-\alpha}{4l}\left(\frac{1+\alpha}{1-\alpha}\right)^l\frac{\Gamma\left(\frac32+l\right)}{\Gamma\left(\frac12-l\right)} 
\end{eqnarray}
and
\begin{eqnarray}
	P_{\frac12}^{-l- \frac l2}(x_r)\approx\frac1{\left(\Gamma\left(\frac32+l\right)\right)}\left(\frac{1-x_r}{1+x_r}\right)^{l/2} \ .
\end{eqnarray}
Substituting these expressions and developing some additional steps we find
\begin{eqnarray}
	U_{Ele}(r)\approx\frac{q^2}{8\alpha r}(\alpha-1)\ln\left(1-r/r_0\right) \ .
\end{eqnarray}

\begin{figure}[tbph]
\begin{center}
\epsfig{figure=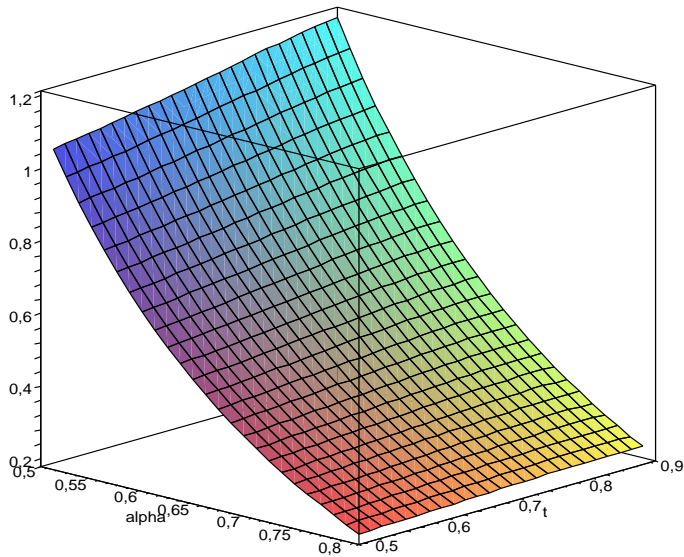,width=10.5cm,height=9.5cm}
\end{center}
\caption{Electrostatic self-energy for a charge inside the monopole core as function of the parameter $\alpha$ and rescaled radial coordinate $t_{in}=r/r_0$.}
\label{fig2}
\end{figure}

Finally in figure \ref{fig2} we present the complete behavior of $r_0U_{Ele}/q^2$ as function of the parameter $\alpha $ and $r/r_0$.

\section{Concluding remarks}
\label{Sect4}
The induced electrostatic self-energy associated with a charged particle placed at rest in the global monopole spacetime, considering it as a point-like topological defect, is divergent at the center of the object \cite{Mello3}. In principle this undesirable result can be avoided by considering a more realistic model for its core. With this objective in mind, in \cite{Mello4} we investigated the influence of a non-trivial structure for the monopole's core on the self-energy considering the flower-pot model for the region inside. For this case the corresponding induced self-energy is expressed in term of two distinct contributions: one coming from the geometry background of the corresponding spacetime, $U_{geom}$, and the other induced by the boundary, $U_{core}$.\footnote{With respect to the region inside, we were able to show that that the self-energy is finite at the  monopole's core.} In order to continue in the same line of investigation, in this paper we decided to revisit the induced electrostatic self-interaction problem, considering at this time a more sophisticated model for the inner structure of the monopole: the ballpoint-pen model. Analogously with happens in the flower-pot model, the renormalized electrostatic self-energy presents two distinct contributions: $U_{geom}$ and $U_{core}$.

As it was mentioned the Introduction, there is no exact solution for the field equations in the region inside the global monopole. The de Sitter model proposed in \cite{HL} describes the geometry for points very close to the monopole's center. As to the model adopted in this paper, like the de Sitter one, it presents a constant scalar curvature. Moreover, we can verify that it is an exact solution for the Einstein equation for the matter field energy-momentum tensor given bellow: 
\begin{eqnarray}
	T^\mu_\nu=\frac{\epsilon^2}{8\pi G\rho_0^2}{\rm diag}(3, \ 1, \ , 1, \,) \ .
\end{eqnarray}

Admitting this model, were able to obtain the renormalized self-interaction in a closed form, and analyzed it for different positions of the charge, and in various asymptotic regions of the parameters. Specifically as to the self-energy at the monopole's center, we found a finite result.   

Combining this result with the previous one found in \cite{Mello4}, we can see that attributing a non-trivial structure for the monopole, the behavior of the self-energy changes drastically, mainly near the boundary and the monopole's center.

Before to finish this section, we would like to mention that, although the contributions due to the geometry background and boundary, are different for the flower-pot and ballpoint-pen models, we observe that they share some similarities: the respective induced self-forces are repulsive with respect to the boundary for $\alpha<1$ and attractive for $\alpha>1$, the self-energy is finite at the monopole's center, and the core-induced contributions are dominant near the boundaries. In fact they are logarithmically divergent, presenting exactly the same expressions for the outer regions.

\section*{Acknowledgment}
DB thanks CAPES for financial support. ERBM thanks CNPq. for partial financial support, FAPESQ-PB/CNPq. (PRONEX) and FAPES-ES/CNPq. (PRONEX). 

\appendix
\section{The renormalization of the Green function $G_0(r',r)$}
\label{app}
The expression for the Green function $G_0(r',r)$ is  
\begin{eqnarray}
\label{g0}
G_0(r',r)=\frac1{4\pi\alpha\sqrt{r'r}}\sum_{l=0}^\infty\frac{(2l+1)(-1)^l\pi}2P_{\frac12}^{l+\frac12}\left(x_{r_>}\right)P_{\frac12}^{-l-\frac12}\left(x_{r_<}\right) \ .
\end{eqnarray}
By using (\ref{Legend}), we can express the term inside the summation by
\begin{eqnarray}
\label{Y}
{\cal{Y}}^{\frac l2+\frac14}F\left( -\frac12, \ \frac32; \ \frac12-l; \ \frac{1-x_{r_<}}2\right)F\left( -\frac12, \ \frac32; \ \frac32+l; \ \frac{1-x_{r_>}}2\right) \ .
\end{eqnarray}
with
\begin{eqnarray}
{\cal{Y}}=\frac{x_{r_<}+1}{x_{r<}-1}\frac{x_{r>}-1}{x_{r>}+1}	 \ ,
\end{eqnarray}
being $x_{r_>}$ and $x_{r_<}$ given in (\ref{X}). In order to observe the divergent contribution of $G_0$, let us analyse (\ref{Y}) in the limit of large value for $l$. The corresponding behavior is
\begin{eqnarray}
	{\cal{Y}}^{\frac l2}+\frac38(x_{r_>}-x_{r_<})\frac{{\cal{Y}}^{\frac l2}}l+O\left(\frac1{l^2}\right) \ .
\end{eqnarray}
The contributions of the first and second terms above in the summation in (\ref{g0}) are, respectively, $\frac1{1-\sqrt{\cal{Y}}}$ and $-\ln(1-\sqrt{\cal{Y}})$. However the second contribution provide a vanishing result in the coincidence limit. Considering  $x_{r_<}\to x_{r_>}$,
\begin{eqnarray}
	{\cal{Y}}\approx 1-2\frac{x_{r_>}-x_{r_<}}{x_{r_<}^2-1}
\end{eqnarray}
and consequently 
\begin{eqnarray}
	1-\sqrt{\cal{Y}}\approx\frac{r_>-r_<}{r_<\sqrt{1-\left(\frac{r_<}{r_0}\right)^2\sin^2\epsilon}} \ .
\end{eqnarray}
Finally the contribution due to this term in (\ref{g0}) is
\begin{eqnarray}
	G_0(r',r)=\frac1{4\pi\alpha}\frac{\sqrt{1-(r_</r_0)^2\sin^2\epsilon}}{r_>-r_<}+ \  ....
\end{eqnarray}
So the renormalized Green function is given by 
\begin{eqnarray}
	G_{0,ren}(r,r)&=&\frac1{4\pi\alpha r}\lim_{r'\to r}\sum_{l=0}^\infty\left[{\cal{Y}}^{\frac l2+\frac14}F\left( -\frac12, \ \frac32; \ \frac12-l; \ \frac{1-x_{r_<}}2\right)\times\right.\nonumber\\
&&\left.F\left( -\frac12, \ \frac32; \ \frac32+l; \ \frac{1-x_{r_>}}2\right)-{\cal{Y}}^{\frac l2}\right] \ ,
\end{eqnarray}
which provide the results (\ref{Gr0}), (\ref{Gr1}) and (\ref{Gr2}).

\end{document}